\documentclass[aps,pra,twocolumn,showpacs,preprintnumbers,amsmath,amssymb,superscriptaddress,10pt,nofootinbib]{revtex4-1}

\usepackage{amsmath,amssymb}
\usepackage{mathrsfs}
\usepackage{bm}
\usepackage[latin1]{inputenc} 
\usepackage{dcolumn}
\usepackage{graphicx, epstopdf}
\usepackage{SIunits}
\usepackage{multirow}
\usepackage{ae}
\usepackage{subfig}
\usepackage{hyperref}
\usepackage{color} 

\newcommand{\diff}[0]{\text{d}}

\newcommand{\re}{\mathrm{Re}}
\newcommand{\im}{\mathrm{Im}}

\newcommand{\be}{\begin{equation}}
\newcommand{\ee}{\end{equation}}
\newcommand{\mueff}{\mu_\text{eff}(\omega)}
\newcommand{\epseff}{\epsilon_\text{eff}(\omega)}

\begin{document}

\title{Relaxed dispersion constraints and metamaterial effective parameters with physical meaning for all frequencies}

\author{Christopher A. Dirdal}
\affiliation{Department of Electronics and Telecommunications, Norwegian University of Science and Technology, NO-7491 Trondheim, Norway}

\author{Tarjei Bondevik}
\affiliation{Department of Electronics and Telecommunications, Norwegian University of Science and Technology, NO-7491 Trondheim, Norway}
\email{}

\author{Johannes Skaar}
\affiliation{Department of Electronics and Telecommunications, Norwegian University of Science and Technology, NO-7491 Trondheim, Norway}
\email{johannes.skaar@ntnu.no}

\date{\today}

\begin{abstract}
Metamaterial effective parameters may exhibit freedom from typical dispersion constraints. For instance, the emergence of a magnetic response in arrays of split-ring resonators for long wavelengths cannot be attained in a passive continuous system obeying the Kramers-Kronig relations. We characterize such freedom by identifying the three possible asymptotes which effective parameters can approach when analytically continued. Apart from their dispersion freedom, we also demonstrate that the effective parameters may be redefined in such a way that they have a certain physical meaning for all frequencies. There exists several possible definitions for the effective permittivity and permeability whereby this is achieved, thereby giving several possible frequency variations for high frequencies, while nevertheless converging to the same dispersion for long wavelengths.
\end{abstract}

\maketitle

\section{Introduction} \label{sec:freedommetamaterials}
The concept of a metamaterial is a powerful one. Complex electromagnetic systems are treated as simple, effectively continuous media with effective homogeneous fields, for which electromagnetic properties unlike those found in any conventional media may emerge. These properties are described by effective parameters $\mu_\text{eff}(\omega)$ and $\epsilon_\text{eff}(\omega)$ which represent the effective permeability and permittivity, respectively, as seen by macroscopic fields in the long wave limit.  Analytic expressions of such parameters have been derived for several systems, including arrays of split-ring cylinders \cite{Pendry1999} in which magnetism is realized from non-magnetic conductors, and L-C loaded transmission lines \cite{Eleftheriades2002} which realize a left handed medium.

In order to examine dispersion properties of metamaterial systems it is natural to consider their effective parameters in relationship with the Kramers-Kronig relations. For the permittivity one has
\begin{subequations}
\begin{align}
\re \ \epsilon(\omega) &= 1 + \frac{2\mathcal{P}}{\pi} \int_{0}^{\infty} \frac{x\im \ \epsilon(x)}{x^2-\omega^2}\diff x \label{eq:KKnormal1} \\
\im \ \epsilon(\omega) &= -\frac{2\omega \mathcal{P}}{\pi}\int_{0}^\infty \frac{\re \ \epsilon(x)-1}{x^2-\omega^2}\diff x, \label{eq:KKnormal2}
\end{align} \label{eq:KKnormal} \\
\end{subequations}where $\mathcal{P}$ represents the principle value. These are derived under the conditions that the permittivity $\epsilon(\omega)\to 1$ as $\omega \to \infty$ and that $\epsilon(\omega)$ must be analytic for $\im \ \omega \geq 0$. On the basis of causality and other physically reasonable assumptions, the permittivity of common media is generally assumed to fulfill these requirements \cite[p.332-333]{jackson_classical_1999}, and hence \eqref{eq:KKnormal}. However, in for example metamaterials consisting of L-C loaded transmission lines this is not necessarily the case. The effective parameters for the particular arrangement of the 1D unit cell displayed in Fig. \ref{fig:unitCell} with series impedance $Z'(\omega)$ and the shunt admittance $Y'(\omega)$ per unit length are \cite{Eleftheriades2002}:

\begin{subequations}
\begin{align}
\epsilon_\text{eff}(\omega) & = -\frac{Y'}{i \omega}. \label{eq:epsTransLine} \\
\mu_\text{eff}(\omega) & = -\frac{Z'}{i \omega} \label{eq:muTransLine}
\end{align} \label{eq:transLine} \\
\end{subequations} If the impedance and admittance of the transmission line are taken to represent common lumped circuit elements such as inductors $Y_L= -i \omega L$, resistors $Y_R= R$ or capacitors $Y_C= -1/i\omega C$, then it is clear from \eqref{eq:epsTransLine} that the asymptotic forms $O(\epseff) = 1, \omega^{-1}, \omega^{-2}$ follow upon analytic continuation for $\omega\to \infty$. The latter two of these cases clearly violate the premisses for \eqref{eq:KKnormal} and hence do not fulfill the Kramers-Kronig relations. 


\begin{figure}[htb]
\begin{center}
\def\svgwidth{0.6\columnwidth}
\begingroup%
  \makeatletter%
  \providecommand\color[2][]{%
    \errmessage{(Inkscape) Color is used for the text in Inkscape, but the package 'color.sty' is not loaded}%
    \renewcommand\color[2][]{}%
  }%
  \providecommand\transparent[1]{%
    \errmessage{(Inkscape) Transparency is used (non-zero) for the text in Inkscape, but the package 'transparent.sty' is not loaded}%
    \renewcommand\transparent[1]{}%
  }%
  \providecommand\rotatebox[2]{#2}%
  \ifx\svgwidth\undefined%
    \setlength{\unitlength}{376.625bp}%
    \ifx\svgscale\undefined%
      \relax%
    \else%
      \setlength{\unitlength}{\unitlength * \real{\svgscale}}%
    \fi%
  \else%
    \setlength{\unitlength}{\svgwidth}%
  \fi%
  \global\let\svgwidth\undefined%
  \global\let\svgscale\undefined%
  \makeatother%
  \begin{picture}(1,0.57524062) 
    \put(0,0){\label{fig:unitCell} \includegraphics[width=\unitlength]{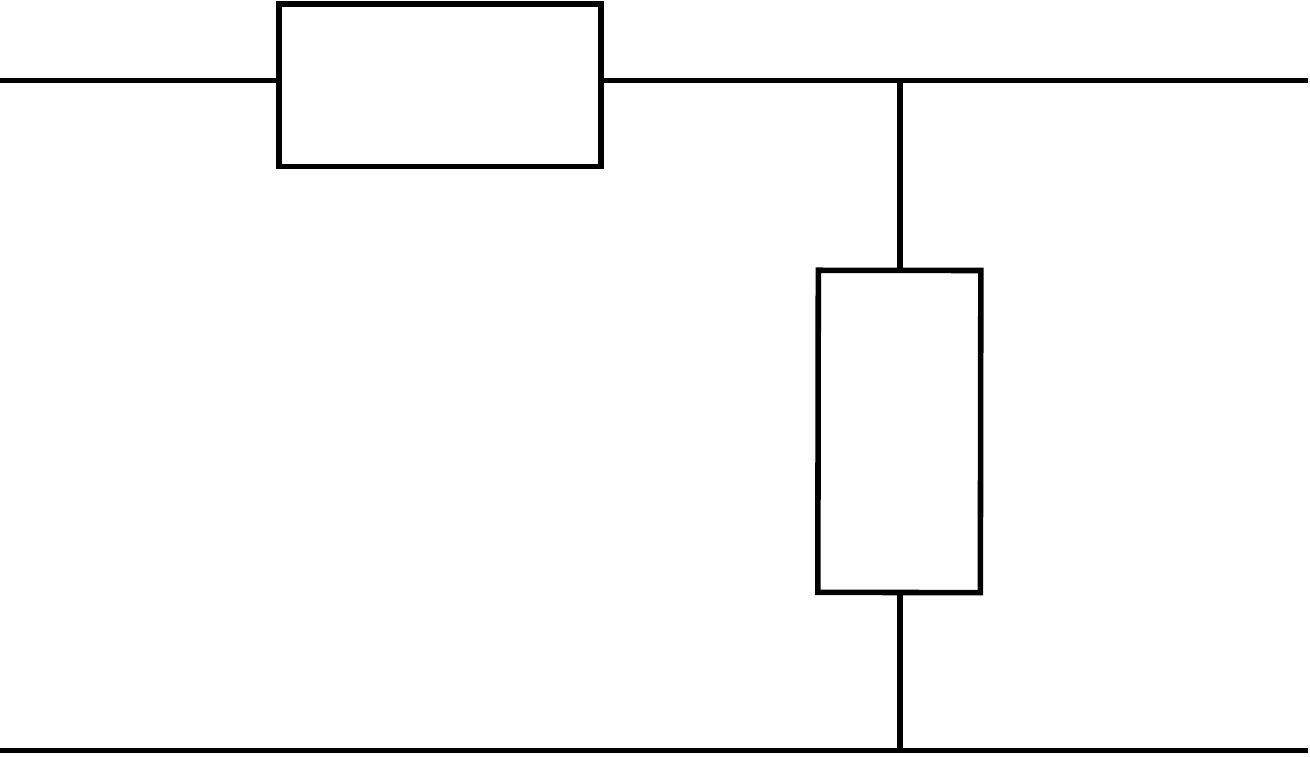}}%
    \put(0.32,0.49){\color[rgb]{0,0,0}\makebox(0,0)[lb]{\smash{$Z'$}}}%
    \put(0.68,0.23){\color[rgb]{0,0,0}\makebox(0,0)[lb]{\smash{$Y'$}}}%
  \end{picture}%
\endgroup 
\end{center}
\caption{Unit cell of a 1D transmission line with series impedance $Z'(\omega)$ and shunt admittance $Y'(\omega)$ per  unit length. }
\label{fig:unitCell}
\end{figure}

Equivalent remarks regarding the effective permeability of  a metamaterial system can be made. It is evident that \eqref{eq:muTransLine} will not obey the Kramers-Kronig relations for impedances $Z_R= R$ and $Z_C= -1/i\omega C$. This is also the case for the effective permeability in another well-known metamaterial system: An array of the split-ring cylinders. The effective parameter as derived by Pendry \cite{Pendry1999} takes the form

\begin{align}
\mu_\text{eff}(\omega) = 1 + \frac{\omega^2 F}{\omega_0^2-\omega^2 -i \omega \Gamma}. \label{eq:Pendrymu}
\end{align} Here $F$ is the volume fraction of the interior of the cylinder in the unit cell, $\omega_0$ is the resonance frequency determined by the cylinder radius and capacitance, and $\Gamma$ is the response width determined by the conductivity and cylinder radius \cite{dirdal2013superpositions}. On account of the deviation from unity of the asymptote of this parameter's analytic continuation

\begin{align}
\mu_\text{eff}(\omega) \to 1 - F, 	\quad \text{as } \omega \to \infty, \label{eq:PendryAsym}
\end{align} the parameter \eqref{eq:Pendrymu} will not obey the Kramers-Kronig relations. As a side remark, the bianisotropy of the split-ring cylinder \cite{PhysRevB.65.144440} is avoided while yielding the same response \eqref{eq:Pendrymu} by a slight modification of the resonator design \cite{Marque2003} \cite[Sec. 16-2]{capolino2009theory}. This ensures that the parameter \eqref{eq:PendryAsym} is local under the constraints discussed further below.

At first glance it might not seem clear that the parameter asymptotes of \eqref{eq:transLine} and \eqref{eq:Pendrymu} being unequal to unity carries any significant practical consequences. However, on account of their analyticity the behavior at all frequencies, even for high frequencies (where the parameters may no longer even be physical, see Sec. \ref{sec:Origin}), is important for how the parameter behaves \textit{in the long wavelength regime}. By violating the assumptions of \eqref{eq:KKnormal} or the K.K. relation for the permeability function, which are generally taken to describe the class of possible dispersions in ordinary media, these metamaterial cases serve as tell-tale signs of some additional dispersion freedom from ordinary dispersion constraints. For instance, it will be shown in the next section that the characteristic occurrence of a magnetic response \eqref{eq:Pendrymu} in an array of split-ring cylinders made from non-magnetic metals is not possible in a passive continuous medium obeying the Kramers-Kronig relations. Hence, questions naturally arise as to what degree dispersion freedom can be attained in general, and what interesting consequences there may be in metamaterials. The purpose of this paper is to characterize this freedom of the effective parameters in relation to the constraints set by the Kramers-Kronig relations, and to explain the origin of their deviation from these. To this end, the following section deduces the possible asymptotic forms which are attainable for passive metamaterial systems. These are then used to identify the space of possible dispersions. Section \ref{sec:Origin} discusses the origin of the parameter freedom in relation to their possible loss of physical meaning for large frequencies and wave numbers $(\omega,k)$. This encourages us to consider alternative definitions of the effective material parameters in Sec. \ref{sec:AltPhysMean}: As will be shown, we may attribute another physical meaning to the parameters which is kept also for wavelengths outside the long wave limit, which nevertheless coincides with the usual meaning of effective permeability or permittivity at low frequencies. Finally, case examples will be presented which illustrate the findings of this article.

This paper considers effective parameters $\mueff$ and $\epseff$ that locally relate the fields of macroscopic electromagnetism. Since spatially dispersive effects have been shown to be a general property of metamaterials \cite[p. 8]{Al`u2011} \cite{Al`u2011a}, local parameters are only possible for long wavelengths, i.e. $kd \ll 1$, where $d$ represents the characteristic size of the relevant non-bianisotropic inclusions, and the wave number $k$ is chosen sufficiently small as to render such effects negligible. Note that this implies that the inclusion parameters then must be chosen suitably to ensure that the relevant response features fall within this region: E.g. for the split-ring cylinder array described by \eqref{eq:Pendrymu} the radius $r$ and capacitance per unit area $C$ must be tailored so that the resonance wave number $k_0=\omega_0/c$ obeys $k_0 r = \sqrt{2/c^2\pi^2\mu_0 C r}\ll 1$. 

For simplicity, the effective parameters are assumed to be scalar throughout the paper. We assume an $\text{e}^{-i\omega t}$ time dependence, meaning that positive imaginary parts of $\epseff$ and $\mueff$ correspond to a dissipation of energy \cite[p. 274]{LandayLifshitz84} (for frequencies $\omega$ in the long wavelength regime).


\section{Characterizing Dispersion Freedom}

\subsection{Asymptotic forms} \label{sec:Charact}
This section will identify the set of possible asymptotic forms which an analytic continuation of the effective parameters can take. When these are known, it is possible to characterize the space of possible dispersions in terms of alternative Kramers-Kronig relations. This space will be shown to be greater than that encompassed by the standard dispersion constraints.

In the doctoral thesis of Otto Brune from 1931 \cite{Brune31} an argument was given which will now be presented here in an adapted form to show that the only three possible asymptotic forms consonant with passivity are 
\begin{align}
O(\mu_\text{eff}),O(\epsilon_\text{eff}) = 1, \omega^{-1}, \ \text{or} \ \omega^{-2}, \quad \text{as} \ \omega \to \infty, \label{eq:ThreeAsymtotes}
\end{align} under the condition that $\omega$ remains in the frequency bandwidth where $\mu_\text{eff}(\omega)$ and $\epsilon_\text{eff}(\omega)$ represent effective permeability and permittivity, respectively. In practical terms, this means that if the long wavelength regime of the physical model underlying $\mu_\text{eff}(\omega)$ or $\epsilon_\text{eff}(\omega)$ is extended indefinitely, for instance by reducing the characteristic sizes of the inclusions, then it follows that $\mu_\text{eff}(\omega)$ and $\epsilon_\text{eff}(\omega)$ must take one of the forms given by \eqref{eq:ThreeAsymtotes}.

For a passive system, the passivity condition (as derived in \cite[\S 80]{LandayLifshitz84}) when extended to the upper complex half-plane for the permeability of a system becomes $\im \ \omega \mu \geq \im \ \omega$ \cite{Liu2013} (under the assumption of no spatial dispersion \cite{QUA:QUA24185}). For our present purposes we observe the necessary condition that $\im \ \omega \mu > 0$. We consider the possible asymptotic forms of $\mu_\text{eff}(\omega)$ in general by assuming
\begin{align}
O ( \omega \mu_\text{eff}) = \omega^n, \quad \text{as }|\omega| \to \infty, \label{eq:asymOmegMu}
\end{align} where $\ n \in \mathbb{Z}$. Then, by expressing $\omega^n  =  |\omega|^n \exp(in\theta)$, one may write 
\begin{align}
O (\im \ \omega \mu_\text{eff}) = |\omega|^n \sin(n\theta), \quad \text{as } |\omega| \to \infty. \label{eq:nCond}
\end{align} Now in order that $\im \ \omega \mu_\text{eff} > 0$ for $\im \ \omega > 0$, it may be seen from \eqref{eq:nCond} that the values of $n$ are restricted to $n \in \{ -1, 0, 1\}$, so as to avoid a sign change for $\theta\in[0,\pi]$. Hence, on the basis of passivity the parameter $\mu_\text{eff}(\omega)$ can only have one of the asymptotic forms \eqref{eq:ThreeAsymtotes}. Equivalent considerations on $\epsilon_\text{eff}(\omega)$ give the same result.

In order to characterize the space of possible dispersions on the basis of the forms \eqref{eq:ThreeAsymtotes}, Kramers-Kronig relations generalized to these can be derived by expressing the effective parameter as  
\begin{align}
\mu_\text{eff}(\omega) \equiv a + \mu_a(\omega)
\end{align} where $a= \lim_{\omega\to \infty}  \mu_\text{eff}(\omega)$, $\mu_a(\omega)$ is square-integrable and $\mu_\text{eff}(\omega)$ is analytic for $\im \ \omega > 0$, and then applying Cauchy's integral theorem. From the symmetry property $\mu_\text{eff}(-\omega)=\mu_\text{eff}^*(\omega^*)$ it follows that the constant $a$ is real. This gives the following expressions

\begin{subequations}
\begin{align}
\re \ \mu_\text{eff}(\omega) &= \ a + \frac{2\mathcal{P}}{\pi} \int_{0}^\infty \frac{x\im \mu_a(x)}{x^2-\omega^2} \diff x \label{eq:GenKK1} \\
\im \ \mu_\text{eff}(\omega) &= -\frac{2 \omega \mathcal{P}}{\pi} \int_{0}^{\infty} \frac{\re \mu_a(x)}{x^2-\omega^2} \label{eq:GenKK2} \diff x.
\end{align} \label{eq:GenKK} \\
\end{subequations} Notice that in the event that $\lim_{\omega\to \infty}\mu_\text{eff}(\omega) = 1$, \eqref{eq:GenKK1} and \eqref{eq:GenKK2} reduce to the conventional Kramers-Kronig relations (where $\mu_a(\omega)$ is set equal to the magnetic susceptibility). Otherwise, they characterize a larger space of dispersions than the conventional Kramers-Kronig relations do. We note also that the K.K. relation suggested for the magnetic permeability by Landau and Lifshitz becomes equal to \eqref{eq:GenKK1} if the parameter $\omega_1$ in \cite[p. 283]{LandayLifshitz84} is set equal to infinity.


The additional freedom present in metamaterial systems can be exemplified by applying the generalized Kramers-Kronig relations \eqref{eq:GenKK} to Pendry's split-ring cylinder metamaterial, where the cylinders are made of a non-magnetic metal, such as aluminum or copper \cite{Pendry1999}. Upon applying a time-varying field, induced current flows on the cylinder surfaces and thereby leads to the emergence of a magnetic response, which is described by $ \mu_\text{eff}(\omega)\neq 1$. At zero frequency, however, only the intrinsic material properties matter, meaning that $\mu_\text{eff}(0)= 1$.  Now, if we were to search for such emergent magnetism in a continuous medium which obeys \eqref{eq:KKnormal}, it turns out that we would necessarily be looking for a gain medium! This is observed from the analogous K.K. relation of \eqref{eq:KKnormal1} for $\mu(\omega)$ with $\omega=0$:

\begin{align}
\re \ \mu(0) = 1 + \frac{2 \mathcal{P}}{\pi} \int_0^\infty \frac{\im \mu(x)}{x} \diff x. \label{eq:NoPassive}
\end{align} The only way to have $\re \ \mu(0) = 1$ for our system is for the integral to equal zero, thereby implying that $\im \ \mu(\omega) < 0$ for some frequencies. This makes the freedom in the metamaterial arrangement explicit: While a continuous medium would need to display gain in order to have this property of emergent magnetism, Pendry's split-ring cylinder metamaterial achieves this while being passive. Equation \eqref{eq:Pendrymu} shows that $\mu_\text{eff}(0)=1$ while $\im \mu_\text{eff}(\omega) \geq 0$ for all positive frequencies. By use of the generalized Kramers-Kronig relation \eqref{eq:GenKK1} where $a=1-F$ according to \eqref{eq:PendryAsym} we find
\begin{align}
\re \ \mu_\text{eff}(0) = 1 - F + \frac{2 \mathcal{P}}{\pi} \int_0^\infty \frac{\im \mu_a(x)}{x} \diff x. \label{eq:GenKKRe}
\end{align} Evidently, one need not assume that $\im \ \mu_\text{eff}(\omega) = \im \mu_a(\omega) < 0$ in order that $\mu_\text{eff}(0) = 1$. 

\subsection{Origin of freedom} \label{sec:Origin}
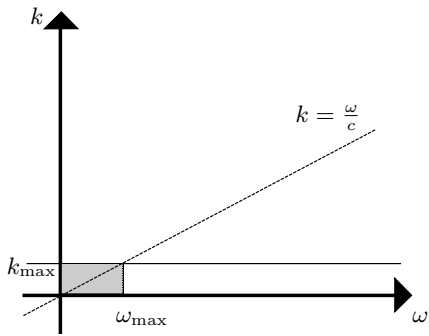
\begin{figure}[htb]
\begin{center}
\def\svgwidth{0.6\columnwidth}
\begingroup%
  \makeatletter%
  \providecommand\color[2][]{%
    \errmessage{(Inkscape) Color is used for the text in Inkscape, but the package 'color.sty' is not loaded}%
    \renewcommand\color[2][]{}%
  }%
  \providecommand\transparent[1]{%
    \errmessage{(Inkscape) Transparency is used (non-zero) for the text in Inkscape, but the package 'transparent.sty' is not loaded}%
    \renewcommand\transparent[1]{}%
  }%
  \providecommand\rotatebox[2]{#2}%
  \ifx\svgwidth\undefined%
    \setlength{\unitlength}{539.48699341bp}%
    \ifx\svgscale\undefined%
      \relax%
    \else%
      \setlength{\unitlength}{\unitlength * \real{\svgscale}}%
    \fi%
  \else%
    \setlength{\unitlength}{\svgwidth}%
  \fi%
  \global\let\svgwidth\undefined%
  \global\let\svgscale\undefined%
  \makeatother%
  \begin{picture}(1,0.9)%
    \put(0,0){\includegraphics[width=\unitlength]{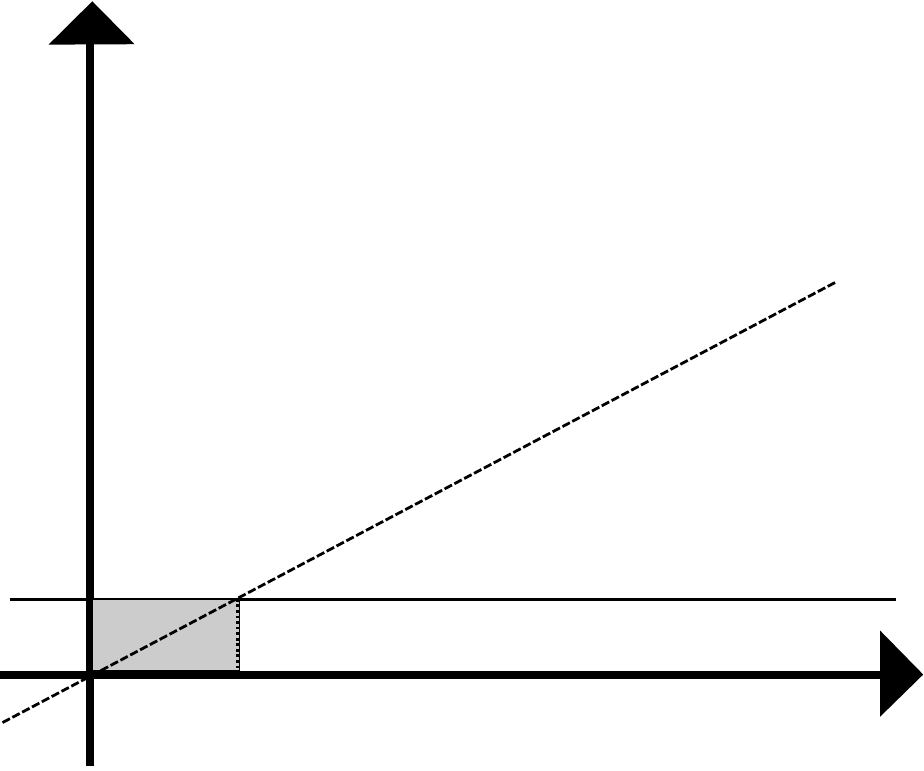}}%
    \put(0.02,0.8){\color[rgb]{0,0,0}\makebox(0,0)[lb]{\smash{$k$}}}%
    \put(1.0,0.03586637){\color[rgb]{0,0,0}\makebox(0,0)[lb]{\smash{$\omega$}}}%
    \put(0.7,0.55){\color[rgb]{0,0,0}\makebox(0,0)[lb]{\smash{$k=\frac{\omega}{c}$}}}%
    \put(0.24,0.03586637){\color[rgb]{0,0,0}\makebox(0,0)[lb]{\smash{$\omega_\text{max}$}}}%
    \put(-0.04,0.16){\color[rgb]{0,0,0}\makebox(0,0)[lb]{\smash{$k_\text{max}$}}}%
  \end{picture}%
\endgroup%
\end{center}
\caption{The subset of $(\omega, k)$ for which the effective parameters $\mu_\text{eff}(\omega)$ and $\epsilon_\text{eff}(\omega)$ represent the effective properties of their corresponding systems according to a suitable homogenization theory (shaded region). The upper eigenmodal frequency $\omega_\text{max}$ has here for simplicity been determined from the intersection between $k=k_\text{max}$ and the line of eigenmodal propagation in vacuum. }
\label{fig:komegSpace}
\end{figure}

So far we have characterized the dispersion freedom stemming from the asymptotic forms \eqref{eq:ThreeAsymtotes} without giving any explanation of their origin. An important observation in this respect, is that the bandwidth of frequencies for which $\mueff$ and $\epseff$ carry the meaning of effective permeability and effective permittivity, respectively, under eigenmodal propagation is usually narrow in comparison to the range of frequencies for which the constituent materials of the metamaterial have an electromagnetic response \cite{LandayLifshitz84,Silveirinha2011}. This is because we only can define local effective parameters in the long wavelength regime $kd \ll 1$ \cite{Al`u2011,Al`u2011a, Silveirinha2007}. When assuming eigenmodal propagation this naturally also implies restrictions upon $\omega$. Furthermore, metamaterial models often use the assumption of quasi-static interactions that may become invalid with increasing $\omega$. Hence the subset in which $\epseff$ and $\mueff$ correspond to the effective permittivity and permeability, respectively, may be classified as $k \leq k_\text{max}$ and $\omega \leq \omega_\text{max}$, where $k$-dependence in $\mu_\text{eff}(\omega)$ and $\epsilon_\text{eff}(\omega)$ is for the sake of illustration assumed negligible below $k_\text{max}$, and $\omega_\text{max}$ is found according to the dispersion-relation or according to quasi-static assumptions (see Fig. \ref{fig:komegSpace}). 


Since for $\omega > \omega_\text{max}$ the effective parameters $\mueff$ and $\epseff$ do not have any direct correspondence with the effective permeability and permittivity of the system, these parameters are not necessarily subjected to ordinary dispersion constraints there. In other words, since the parameters cease to represent the effective properties of the medium they need not necessarily comply with the usual Kramers-Kronig relations or even passivity requirements. Since the parameters are analytic, this freedom at large frequencies also leads to dispersion freedom at low frequencies for which interesting physical consequences may occur. This is the origin of the freedom that has been characterized in the previous section. 


As a side remark, one should note that source-driven systems do not in general need to adhere to the dispersion relation: Any combination of $\omega$ and $k$ is in principle possible provided one is able to supply the necessary arrangement of sources  within the medium \cite{Al`u2011, Silveirinha2007}. Physical $\mu_\text{eff}(\omega)$ and $\epsilon_\text{eff}(\omega)$ may therefore in principle be defined for all $\omega$ if the medium is source driven to keep $k < k_\text{max}$, and the parameters are obtained from an appropriate homogenization procedure \cite{Al`u2011}. 





\section{Parameter definitions that give alternative physical meaning for all $(\omega,k)$} \label{sec:AltPhysMean}

Following the discussions in Sec. \ref{sec:Origin}, it may be argued that the form of the frequency variation in  $\mueff$ given by \eqref{eq:Pendrymu} for $\omega > \omega_\text{max}$ is in some sense arbitrary. What if we therefore were to alter the mathematical expression of $\mueff$ in such a way that for $\omega>\omega_\text{max}$ one instead has $\mueff \to 1$ as $\omega\to \infty$, while allowing it to (approximately) keep the response shape \eqref{eq:Pendrymu} for $\omega < \omega_\text{max}$? According to the discussion in Sec. \ref{sec:Charact} regarding \eqref{eq:NoPassive}, such a modified $\mueff$ would obey the conventional K.K. relations and have $\im \mueff < 0$ in some region where $\omega > \omega_\text{max}$.  Thereby, instead of observing dispersion freedom as a violation of the conventional K.K. relations, it is now seen in terms of having  $\im \mueff < 0$ for a passive system. This therefore constitutes an alternative way of visualizing the same dispersion freedom that was discussed in Sec. \ref{sec:Charact}.  We shall examine this further in the first of the following case examples. 

A second interesting point may be observed from the above illustration. It is possible to construct for the effective parameters $\epseff$ and $\mueff$ of a given metamaterial system a number of different frequency variations for large $\omega$ and $k$. On this note, we shall in the following show that it is possible and arguably preferable to redefine the parameters $\epseff$ and $\mueff$ in such a way that their frequency variations have a clear, physical meaning for all $(\omega, k)$. Since the concepts of local permittivity and permeability functions are not extendable to the entire space $(\omega, k)$, this necessarily implies giving $\epseff$ and $\mueff$ \emph{other physical meanings} that nevertheless coincide with that of effective permittivity and permeability for $\omega\leq \omega_\text{max}$ and $k \leq k_\text{max}$. A general procedure by which this is ensured shall now be presented in the case of the effective permeability function $\mueff$ (an analogous approach may be employed in the case of $\epseff$), before moving on to two case examples. 

We select for the parameter $\mu_\text{eff}(\omega)$ of an arbitrary metamaterial, the following definition
\begin{align}
\mathbf{B}_\text{av}^\text{d}(\omega)= \mu_0 \mueff \mathbf{H}_\text{av}^\text{d}(\omega) \label{eq:GenDefMu}
\end{align} where 
\begin{subequations}
\begin{align}
\mathbf{B}^\text{d}_\text{av} &= \mu_0 \mathbf{\bar{H}} + \frac{1}{V}\int_V \mathbf{M}(\mathbf{r})\diff^3 \mathbf{r} \label{eq:avdipB} \\
\mathbf{H}^\text{d}_\text{av} &= \mathbf{\bar{H}} + \frac{i \omega}{V}\int_V \frac{\mathbf{r} \times \mathbf{P}(\mathbf{r})}{2} \diff^3 \mathbf{r} \label{eq:avdipH} \\
\mathbf{\bar{H}} &= \frac{1}{V} \int_V \mathbf{H}(\mathbf{r}) \text{e}^{-i \mathbf{k}\cdot \mathbf{r}} \diff^3 \mathbf{r}. \label{eq:avH}
\end{align} \label{eq:AverFields} \\
\end{subequations} Here $\mathbf{B}_\text{av}^\text{d}(\omega)$ and $\mathbf{H}_\text{av}^\text{d}(\omega)$ generally represent the dominant terms of the averaged field expansions according to the homogenization approach outlined in \cite{Al`u2011}, $\mathbf{M}(\mathbf{r})$ and $\mathbf{P}(\mathbf{r})$ represent the microscopic induced magnetization and electrical polarization, respectively, and $V$ represents the volume of the unit cell. Even though the physical interpretation of the quantities \eqref{eq:AverFields} themselves are not intuitive for $\omega > \omega_\text{max}$ and $k>k_\text{max}$, the concrete physical meaning of $\mu_\text{eff}(\omega)$ by \eqref{eq:GenDefMu} is kept for all $(\omega,k)$. This concrete physical meaning is nothing more than a relationship between $\mathbf{B}_\text{av}^\text{d}(\omega)$ and $\mathbf{H}_\text{av}^\text{d}(\omega)$, which is itself without any clear physical interpretation for $\omega > \omega_\text{max}$ and $k > k_\text{max}$. Nevertheless it is helpful to know exactly what the effective parameter represents there, especially when the parameters display somewhat odd properties such as those discussed in Sec. \ref{sec:freedommetamaterials} and Sec. \ref{sec:Charact}. Furthermore, by virtue of \eqref{eq:GenDefMu} the K.K relations of the redefined parameter will now relate an unambiguous quantity. Although the parameter $\mu_\text{eff}(\omega)$ no longer represents the effective permeability of the system \emph{in general}, its meaning nevertheless coincides with the \emph{particular} meaning of effective permeability in the long wave limit $k \to 0$, according to the homogenization procedure \cite{Al`u2011}. 

The above procedure by \eqref{eq:GenDefMu}-\eqref{eq:AverFields} is not unique in its ability to attribute an alternative physical meaning to the parameter $\mueff$ for all frequencies, however the procedure is quite general.

\subsection{Split-ring cylinders} \label{sec:SRC}
Here the definition \eqref{eq:GenDefMu} will be used to give the parameter $\mueff$ of an array of split-ring cylinders the alternative physical meaning discussed above for all $(\omega,k)$. As shall be shown, a consequence of this redefinition under the assumption of eigenmodal propagation is that the analytic continuation of the parameter $\mueff$ will approach unity, instead of $1-F$ as is the case for \eqref{eq:Pendrymu}. Thus, the redefinition of the parameter by \eqref{eq:GenDefMu} here succeeds in redefining the parameter in the manner discussed at the introduction of this section. Equation \eqref{eq:GenDefMu} will therefore give a $\mueff$ with a frequency variation which obeys the conventional K.K. relations, and will display $\im \ \mueff < 0$ for some frequencies while still representing a passive medium.

In the split-ring cylinder metamaterial, we consider a polarization where $\mathbf{B}^\text{d}_\text{av}$ and $\mathbf{H}^\text{d}_\text{av}$ are parallel to the cylinder axes. Thus $\mueff$ is a scalar which according to \eqref{eq:GenDefMu} may be expressed as the ratio of the field quantities
\be
\mu_\text{eff}(\omega) = \frac{B_\text{av}^\text{d}(\omega)}{\mu_0 H_\text{av}^\text{d}(\omega)}  = \frac{1}{1- M_\text{net}/\bar{H}}. \label{eq:GenDefMu2}
\ee Here we have solved for \eqref{eq:avdipB}-\eqref{eq:avH} and inserted them into \eqref{eq:GenDefMu} under the assumption that the cylinders have thin walls of a non-magnetic metal, and defined 
\begin{align}
M_\text{net}(\omega)= \frac{F}{2\pi} \int_0^{2\pi} J(r, \phi)|_{r=R} \diff \phi,
\end{align} which represents the net magnetization density. The constant $F$ represents the volume fraction of the cylinder in a unit cell and $J$ represents the current per cylinder length flowing on the cylinders with radius $R$. Under the assumption of eigenmodal propagation ($k \to\omega / c$ as $\omega \to \infty$) one has that $M_\text{net}/\bar{H}\to 0$ as $\omega \to \infty$ on account of the conductivity of a metal tending to zero here. From \eqref{eq:GenDefMu2} it therefore becomes clear that $\mueff \to 1$ as $\omega\to \infty$, meaning that it obeys the conventional K.K.-relations

\begin{subequations}
\begin{align}
\re \ \mu_\text{eff}(\omega) &= 1 + \frac{2\mathcal{P}}{\pi} \int_{0}^{\infty} \frac{x\im \ \mu_\text{eff}(x)}{x^2-\omega^2}\diff x \label{eq:KKnormal1mueff} \\
\im \ \mu_\text{eff}(\omega) &= -\frac{2\omega \mathcal{P}}{\pi}\int_{0}^\infty \frac{\re \ \mu_\text{eff}(x)-1}{x^2-\omega^2}\diff x, \label{eq:KKnormal2mueff}
\end{align} \label{eq:KKnormalmueff} \\
\end{subequations} provided that the resulting $\mueff$ is analytic. 

By \eqref{eq:KKnormal1mueff} evaluated at $\omega=0$ it is clear that one must have  $\im \ \mueff < 0$ for some frequencies when demanding $\mu_\text{eff}(0)=1$, as discussed in Sec. \ref{sec:Charact}. Thus one observes that the parameter $\mueff$ found from \eqref{eq:GenDefMu2} by solving for $M_\text{net}(\omega)$ is quite different from \eqref{eq:Pendrymu}. Both expressions \eqref{eq:Pendrymu} and \eqref{eq:GenDefMu2} must nevertheless coincide in the limit $\omega \to 0$, since they both arise from valid homogenization procedures. In fact \eqref{eq:GenDefMu2} becomes identical to \eqref{eq:Pendrymu} if $M_\text{net}(\omega)$ is solved under the assumption of quasistatic interactions. As a side remark, herein lies the explanation of why different frequency variations are permitted for the effective parameters of a single metamaterial system, despite the existence of a unique continuation of an analytic parameter: There exists a variety of possible definitions of $\mu_\text{eff}(\omega)$ which attain the particular meaning of the effective permeability in the limit $k \to 0$. Since these will generally be approximations of the effective permeability in the long wavelength regime, each definition of $\mu_\text{eff}(\omega)$ may exhibit small, perhaps negligible, deviations from one another here. It follows that each of their respective continuations to higher frequencies, though unique, may nevertheless deviate significantly from one another.




We have demonstrated both that the parameter $\mueff$ for a split-ring cylinder metamaterial may take different frequency variations for large frequencies and obey different K.K. relations, without actually finding the net magnetization density $M_\text{net}(\omega)$ in \eqref{eq:GenDefMu2}. Solving $M_\text{net}(\omega)$ for all frequencies poses difficulties beyond the scope of this article. However, in order to illustrate the findings discussed above, we shall now pursue a qualitative and slightly arbitrary approach, while referring the interested reader to the rigorous method presented for determining $\mueff$ by \eqref{eq:GenDefMu} and \eqref{eq:AverFields} in \cite{Al`u2011}. If the interaction between the cylinders is modeled quasi-statically, straightforward application of Faraday's law \cite{Pendry1999, dirdal2013superpositions} gives for the magnetization density
\begin{align}
M_\text{net}^\text{q.s.}(\omega) = \frac{\omega^2\bar{H}F}{\omega_0^2 -\omega^2(1-F) -i \omega \Gamma}, \label{eq:JnetquasiStatic}
\end{align} where $\omega_0$ is the resonance frequency determined by the cylinder radius and capacitance, and $\Gamma$ is the response width determined by the conductivity and cylinder radius. The asymptote of \eqref{eq:JnetquasiStatic} is erroneous in that it does not approach zero for infinite frequencies. Rather than calculate the exact, dynamic $M_\text{net}(\omega)$ we instead correct the asymptote of the quasistatic solution by multiplying it with a Lorentzian response function with resonance outside the quasi-static limit $\omega_\text{r} \gg \omega_0$ and width $\Gamma_\text{m}$, according to

\begin{align}
M_\text{net}^\text{arb}(\omega) = M_\text{net}^\text{q.s.}(\omega) \frac{\omega_\text{r}^2}{\omega_\text{r}^2-\omega^2 -i \omega \Gamma_\text{m}}, \label{eq:ChangedJ}
\end{align} which therefore ensures that the corrected function is analytic. Although this choice is arbitrary, it nevertheless approximates the correct solution for $M_\text{net}(\omega)$ in the quasi-static regime and vacuum limit. Inserting \eqref{eq:ChangedJ} for $M_\text{net}(\omega)$ in \eqref{eq:GenDefMu2} thus yields a $\mueff$ which can be verified to be analytic, and which obeys \eqref{eq:KKnormalmueff}.

\begin{figure}[htb]
\centering
\subfloat[The parameter $\mu_\text{eff}(\omega)$ for the split ring cylinder arrangement in \cite{Pendry1999}, for a quasi-static model of the fields \eqref{eq:Pendrymu} where the magnetization is given by \eqref{eq:JnetquasiStatic}. This gives the asymptote $\mueff \to 1-F$ when analytically continued to high frequencies, assuming a filling factor $F=0.5$. Hence $\mu_\text{eff}(\omega)$ fulfills the generalized Kramers-Kronig relations \eqref{eq:GenKK} with $a=0.5$.]{\label{fig:PendryResp}\includegraphics[width=0.45\textwidth]{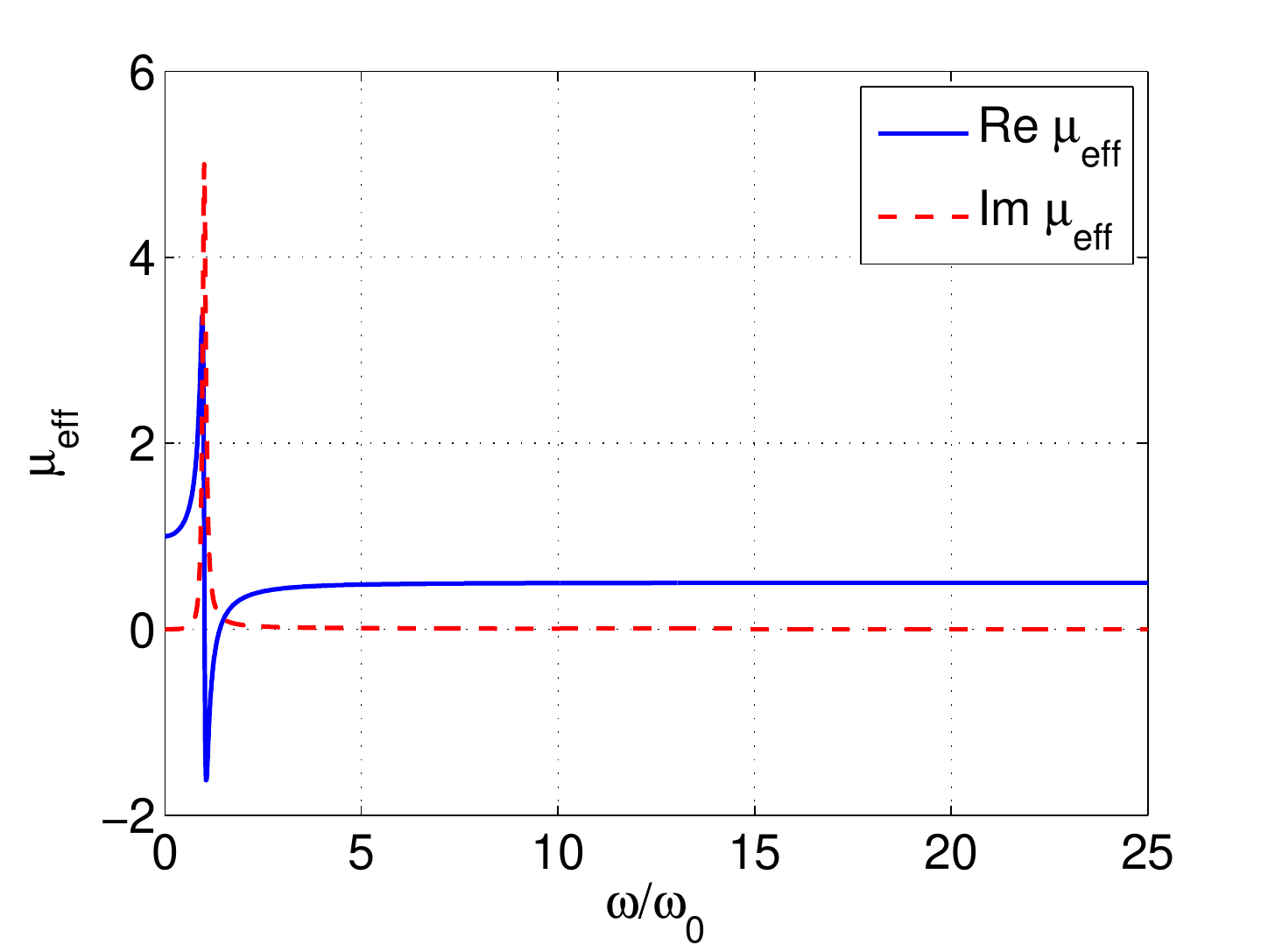}} \\
\subfloat[The parameter $\mu_\text{eff}(\omega)$ for the same system, now using a modified magnetization density model $M_\text{net}^\text{arb}(\omega)$ according to \eqref{eq:ChangedJ}. This results in $\mu_\text{eff}(\omega) \to 1$ as $\omega \to \infty$, as we expect for a dynamic model under eigenmodal propagation. Hence $\mu_\text{eff}(\omega)$ now fulfills the conventional Kramers-Kronig relations \eqref{eq:KKnormal}, and as a result displays $\im \ \mu_\text{eff}(\omega) < 0$ for some frequencies, as implied by \eqref{eq:KKnormalmueff}.]{\label{fig:ExtrapoPendryResp}\includegraphics[width=0.45\textwidth]{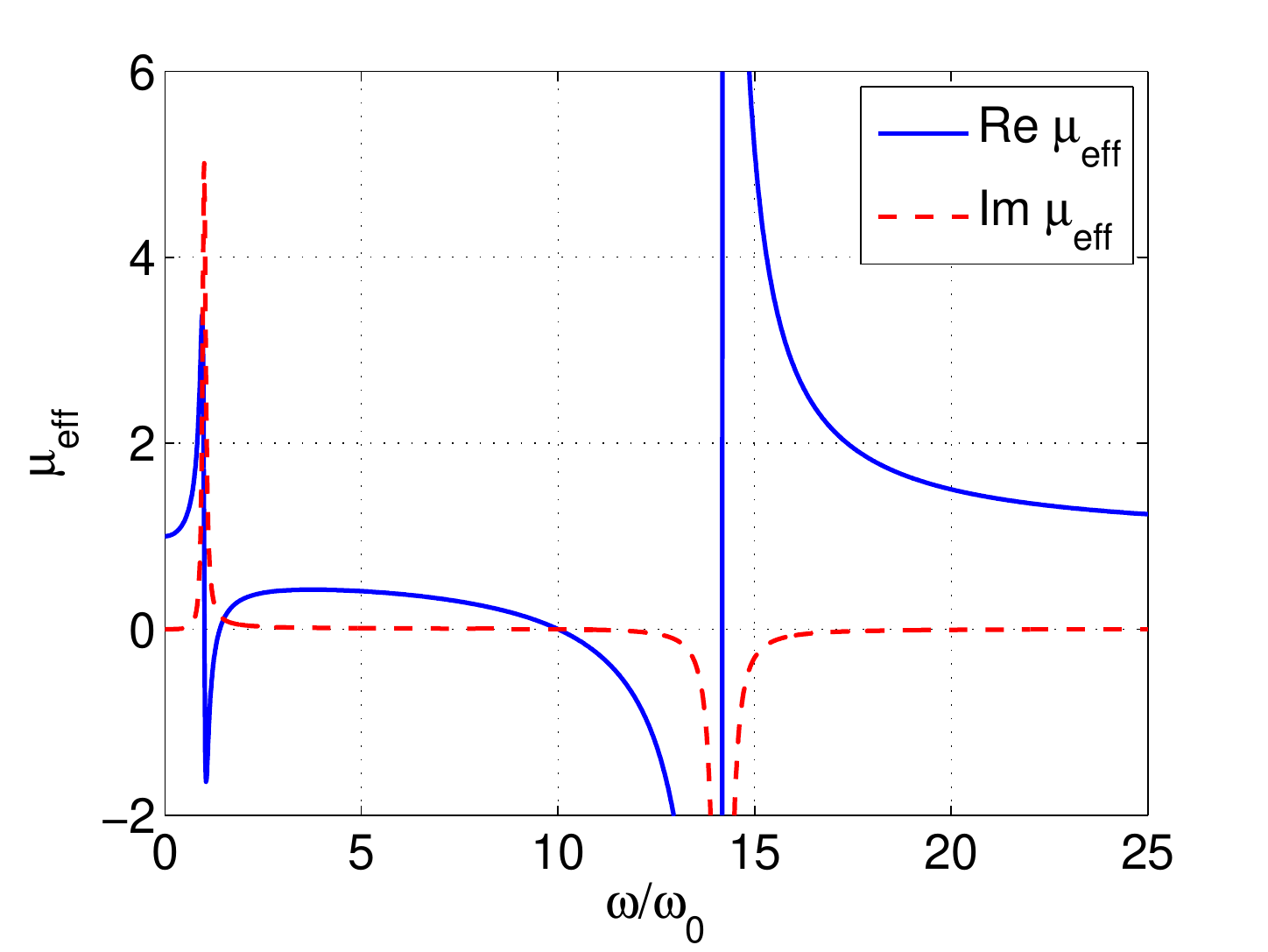}} 
\caption{}
\label{fig:PendryResponses}
\end{figure}

Figure \ref{fig:PendryResponses} compares the frequency variations of $\mueff$ by \eqref{eq:Pendrymu} in Fig. \ref{fig:PendryResp} and by \eqref{eq:GenDefMu2} in Fig. \ref{fig:ExtrapoPendryResp}. One observes that both frequency variations are essentially equal for low frequencies, while differing for larger frequencies with different asymptotes towards the vacuum limit. For the intermediate regime one observes a negative value of $\im \ \mu_\text{eff}(\omega)$ in Fig. \ref{fig:ExtrapoPendryResp} as predicted by \eqref{eq:KKnormalmueff}, although the particular plot here does not represent the actual definition of $\mu_\text{eff}(\omega)$ by \eqref{eq:GenDefMu2} owing to the arbitrary choice we have made for $M_\text{net}(\omega)$ by \eqref{eq:ChangedJ}. The definition of $\mu_\text{eff}(\omega)$ according to \eqref{eq:GenDefMu2} allows us to make sense of what a negative $\im \ \mu_\text{eff}(\omega)$ means: Since it does not occur in the long wave region, it need not be interpreted to indicate gain (which would be nonsensical for a passive system), but represents a phase difference greater than $\pi$ between the quantities $B_\text{av}^\text{d}(\omega)$ and $H_\text{av}^\text{d}(\omega)$. 

Figure \ref{fig:PendryResponses} is interesting in reference to the discussion in Sec. \ref{sec:Charact} regarding the different asymptotic forms \eqref{eq:ThreeAsymtotes} that can be present in a metamaterial. The three asymptotic forms discussed there encompass all possible dispersions for metamaterials \emph{in the long wave limit}. Contrary to the first impression one might get from Fig. \ref{fig:PendryResponses}, the parameters as given by \eqref{eq:Pendrymu} and \eqref{eq:GenDefMu2} in this sense have the same asymptotic form $\mueff \to 1-F$. Put more precisely, if one assumes that the long wavelength regime is extended indefinitely by e.g. reducing inclusion sizes so that $\omega_\text{max} \to \infty$ before we let $\omega\to \infty$, both \eqref{eq:Pendrymu} and \eqref{eq:GenDefMu2} would give $\mueff\to 1-F$. Hence, despite the different asymptotes in Fig. \ref{fig:PendryResp} and Fig. \ref{fig:ExtrapoPendryResp} for fixed $\omega_\text{max}$, both parameters nevertheless carry the same dispersion freedom as characterized by \eqref{eq:ThreeAsymtotes} which leads to the property of emergent magnetism discussed in Sec. \ref{sec:Charact}. This freedom is however manifested differently: In Fig. \ref{fig:PendryResp} one observes a frequency variation which does not obey the conventional K.K. relations, while in Fig. \ref{fig:ExtrapoPendryResp} one observes $\im \mueff < 0$ even though the metamaterial is passive.

\subsection{1D Bragg stack} \label{sec:1DBragg}


This section presents a simpler approach by which an effective parameter may be redefined to have alternative physical meaning for all frequencies. We shall consider a 1D Bragg stack under eigenmodal propagation, for which the solution is straightforward. We choose as our parameter $n_\text{eff}(\omega)$, which will give the effective refractive index in the long wave limit. As will be shown, this parameter will depend on the $z$-coordinate along the axis of periodicity for frequencies outside the long wavelength regime. Hence, it will be possible to have a large number of different effective frequency variations for $n_\text{eff}(\omega)$ at large frequencies, each corresponding to different values of $z$, which all nevertheless converge to the same dispersion in the long wave limit. Some of these may exhibit $\im \ n_\text{eff}(\omega)<0$ for some $\omega$, while others will have only positive imaginary parts.

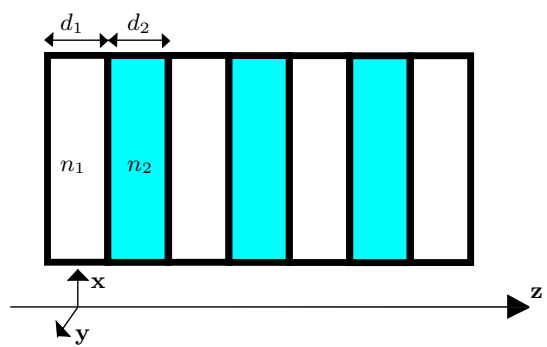
\begin{figure}[htb]
\def\svgwidth{0.8\columnwidth}
\begingroup%
  \makeatletter%
  \providecommand\color[2][]{%
    \errmessage{(Inkscape) Color is used for the text in Inkscape, but the package 'color.sty' is not loaded}%
    \renewcommand\color[2][]{}%
  }%
  \providecommand\transparent[1]{%
    \errmessage{(Inkscape) Transparency is used (non-zero) for the text in Inkscape, but the package 'transparent.sty' is not loaded}%
    \renewcommand\transparent[1]{}%
  }%
  \providecommand\rotatebox[2]{#2}%
  \ifx\svgwidth\undefined%
    \setlength{\unitlength}{302.4281472bp}%
    \ifx\svgscale\undefined%
      \relax%
    \else%
      \setlength{\unitlength}{\unitlength * \real{\svgscale}}%
    \fi%
  \else%
    \setlength{\unitlength}{\svgwidth}%
  \fi%
  \global\let\svgwidth\undefined%
  \global\let\svgscale\undefined%
  \makeatother%
  \begin{picture}(1,0.67871419)%
    \put(0,0){\includegraphics[width=\unitlength]{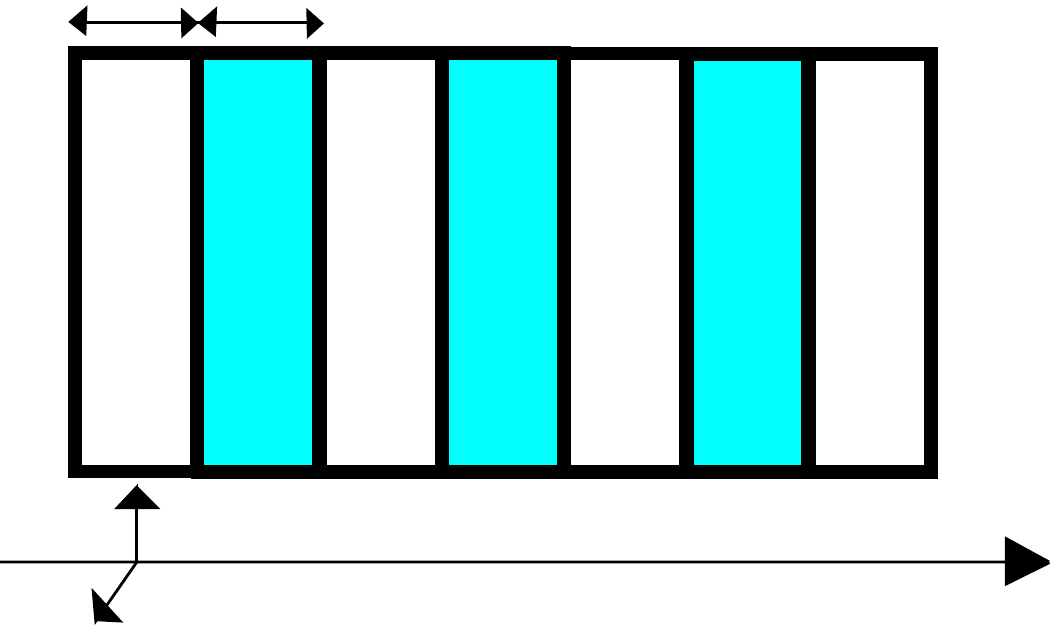}}%
    \put(0.09566012,0.59599826){\color[rgb]{0,0,0}\makebox(0,0)[lb]{\smash{$d_1$}}}%
    \put(0.225,0.59599826){\color[rgb]{0,0,0}\makebox(0,0)[lb]{\smash{$d_2$}}}%
    \put(0.09566012,0.325){\color[rgb]{0,0,0}\makebox(0,0)[lb]{\smash{$n_1$}}}%
    \put(0.225,0.325){\color[rgb]{0,0,0}\makebox(0,0)[lb]{\smash{$n_2$}}}%
    \put(0.155,0.10){\color[rgb]{0,0,0}\makebox(0,0)[lb]{\smash{$\mathbf{x}$}}}%
    \put(0.125,0.0){\color[rgb]{0,0,0}\makebox(0,0)[lb]{\smash{$\mathbf{y}$}}}%
    \put(1.0,0.08){\color[rgb]{0,0,0}\makebox(0,0)[lb]{\smash{$\mathbf{z}$}}}%
  \end{picture}%
\endgroup%
\caption{1D photonic crystal with alternating dielectric layers where $n_1 < n_2$.}
\label{fig:1DPhotonicCryst}
\end{figure}

\begin{figure}[htb]
\subfloat[The parameter $n_{z,\text{eff}}(\omega)$ as given by \eqref{eq:neffbytransfer} where $z= 5 (d_1+d_2)$, i.e. $z$ is chosen within a lower index layer with index $n_1=1$. The higher index layer has $n_2=10 + 0.1i$.]{\label{fig:1DNoGain}\includegraphics[width=0.45\textwidth]{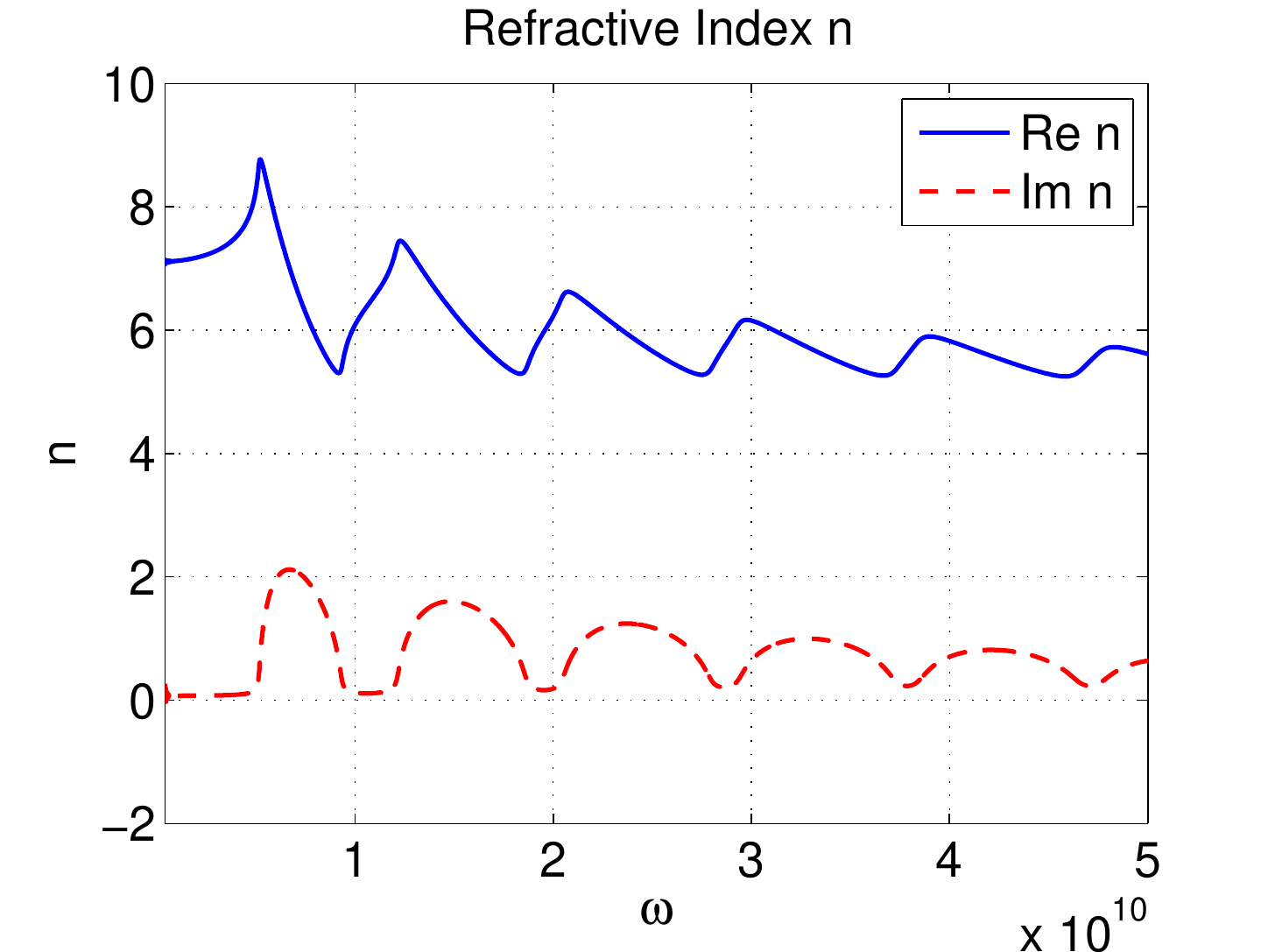}} \\
\subfloat[The parameter $n_{z,\text{eff}}(\omega)$ as given by \eqref{eq:neffbytransfer} for the same photonic crystal, but where $z= 5.5 (d_1+d_2)$, i.e. $z$ is chosen within a high index layer with index $n_2$. Notice that $\im n_\text{eff} < 0$ for a small bandwidth even though the photonic crystal is passive. This occurs as a result of a transfer function $G>1$ in \eqref{eq:neffbytransfer} through the accumulation of field in the high index layer.]{\label{fig:1DGain}\includegraphics[width=0.45\textwidth]{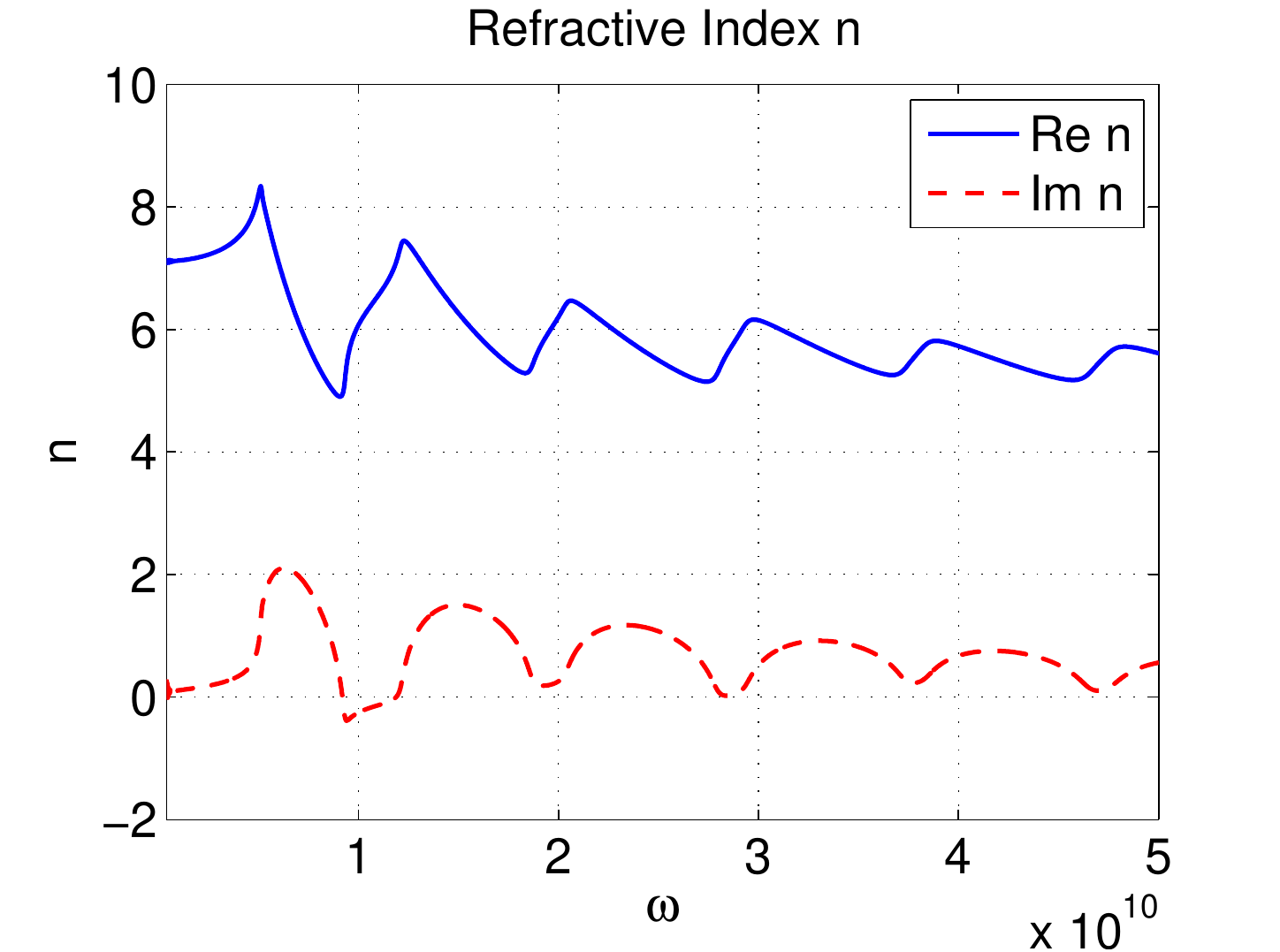}} 
\caption{}
\label{fig:1DCrystal}
\end{figure}

Consider the 1D photonic crystal as shown in Fig. \ref{fig:1DPhotonicCryst}. We imagine placing a current sheet at $z=0$ in the layer of lowest refractive index $n_1$ in which the current alternates as $\mathbf{J}_s=J_0 \mathbf{\hat{x}}$ (with the harmonic time variation suppressed). Noting that the resulting fields must be continuous over the interfaces, it follows that for $\omega \to 0$ the magnetic field approaches that of an effective continuous medium


\begin{align}
\mathbf{H}=-\frac{J_0}{2}\exp(in_\text{eff}\omega z /c )\mathbf{\hat{y}} \quad \text{for } 0 < z < \frac{d_1}{2},
\end{align} in terms of an effective index of refraction $n_\text{eff}$. Therefore, as $\omega \to 0$ the transfer function 
\be
G\equiv \frac{H(z)}{H(0^{+})}, \label{eq:Transferfunc}
\ee approaches that of a continuous medium:
\begin{align}
G= \exp(i n_\text{eff} \frac{\omega}{c}z). \label{eq:TransfCont}
\end{align} Motivated by the form of \eqref{eq:TransfCont} we obtain our definition:
\begin{align}
n_{z,\text{eff}}(\omega) \equiv-\frac{ic}{\omega z}\ln G \label{eq:neffbytransfer}
\end{align} where $G$ is given by \eqref{eq:Transferfunc}. Now our parameter $n_{z,\text{eff}}(\omega)$ is in general a quantity proportional to the logarithm of the transfer function in the medium; a quantity that has physical meaning for all frequencies. For low frequencies, this physical meaning coincides with that of the effective refractive index. The subscript $z$ indicates that the transfer function evaluated at each value of $z$ yields a different frequency variation. All of these converge to the same $n_\text{eff}(0)$ as $\omega \to 0$, for which the analytic value may be calculated to be
\begin{align}
n_\text{eff}(0)= \sqrt{\frac{n_1^2 d_1 + n_2^2d_2}{d_1+d_2}}. \label{eq:neffzero}
\end{align} Provided the transfer function $G$ does not have zeros in the upper complex half-plane, our parameter $n_{z,\text{eff}}(\omega)$ is analytic there while having the asymptote $n_{z,\text{eff}}(\omega) \to 1$, and therefore obeys the conventional Kramers-Kronig relations.

Figure \ref{fig:1DCrystal} displays two possible variations as given by \eqref{eq:neffbytransfer} where $z$ is chosen within a layer of the low index $n_1=1$ in Fig. \ref{fig:1DNoGain} and where $z$ is chosen within a layer of the higher index $n_2=10 + 0.1i$ in Fig. \ref{fig:1DGain}. The transfer function \eqref{eq:Transferfunc} has been calculated by use of the relevant transfer matrices and boundary conditions. The indexes $n_1$ and $n_2$ have been assumed constant with respect to frequency for simplicity. This means that for $\omega \to \infty$ one has $n_{z,\text{eff}}(\omega)\to n_\infty \neq 1$. As a side remark, we note that a similar effective parameter definition for the effective refractive index with a similar plot as that in Fig. \ref{fig:1DNoGain} is proposed in \cite{Liu2013}. Figures \ref{fig:1DNoGain} and \ref{fig:1DGain} give slightly different variations: Fig. \ref{fig:1DNoGain} has $\im \ n_{z,\text{eff}}(\omega) \geq 0$ for all $\omega$, while Fig. \ref{fig:1DGain} displays $\im \ n_{z,\text{eff}}(\omega) < 0$ for a small bandwidth. This demonstrates that the parameter may be redefined to give frequency variations both with and without $\im \ n_\text{eff} (\omega) < 0$. This corresponds to the multiple ways in which one may define a physical quantity that is meaningful for all frequencies, which at the same time approximates the refractive index of the medium for small frequencies.

In light of the physical definition of $n_{z,\text{eff}}(\omega)$, we notice that the negative imaginary part in Fig. \ref{fig:1DGain} has nothing to do with gain: A negative value of $\im \ n_{z,\text{eff}}(\omega)$ in \eqref{eq:neffbytransfer} corresponds to a transfer function greater than unity. This occurs as a result of a Fabry-Perot interference occurring between the low and high index layers, leading to a local accumulation of field in the high index layer where the transfer function is evaluated.


\section*{Conclusions} 
In this article we have identified examples of metamaterial effective parameters that do not follow conventional dispersion constraints, as represented by the Kramers-Kronig relations. This freedom in dispersion has been characterized through the identification of the three asymptotes which the analytic continuation of the effective parameters $\mu_\text{eff}(\omega)$ and $\epsilon_\text{eff}(\omega)$ may approach. The space of possible dispersions in metamaterials have been identified through generalizing the Kramers-Kronig relations for the three possible asymptotes. The possibility of redefining metamaterial parameters so as to achieve a certain physical meaning for all frequencies 
has also been presented. Such an approach may be both aesthetically and practically motivated. Regarding the former the Kramers-Kronig relations of the redefined parameters generally relate an unambiguous physical quantity. Regarding the latter, it can be useful to have a clear definition of what the effective parameters represent for large frequencies, as in the two case examples considered where the occurrence of negative imaginary parts in the effective parameters of two passive metamaterials have been given an intuitive physical explanation.

\end{document}